\documentstyle[twocolumn,aps]{revtex}
\begin{document}

\draft

\title{Low-Lying Excitations of Quantum Hall Droplets}

\author{J.H.Oaknin$^1$,  L.Mart\'{\i}n-Moreno$^2$,
J.J.Palacios$^1$\cite{address} and C.Tejedor$^1$}
\address{$^1$Departamento de F\'{\i}sica de la Materia Condensada,
Universidad Aut\'onoma de Madrid, Cantoblanco, 28049, Madrid, Spain.}

\address{$^2$Instituto de Ciencia de Materiales (CSIC),
Universidad Aut\'onoma de Madrid, Cantoblanco, 28049, Madrid, Spain.}

\date{\today}

\maketitle

\begin{abstract}
We present the adequate circular representation of charge
density excitations in Quantum Hall droplets at filling factor 1.
A novel set of operators describes magnetoexcitons
with well defined both total and center-of-mass angular momentum.
The accuracy of this description is shown by the high overlap of the
magnetoexcitons with the exact eigenstates of a quantum dot in a
magnetic field obtained from numerical calculations up to 20
electrons. From such magnetoexcitons we get a good understanding of total
energies
and spectral functions.
\end{abstract}
PACS number: 73.40.Hm

\begin{center}********************************** \end{center}

The electronic structure of quantum Hall droplets (QHD) has been studied
extensively
\cite{3,trugman,4,Wen,Stone,5,Stone2,7,9,10,Marsili,MacDonald,Pitaevskii,nos},
both due to its relevance in the analysis of quasi-two
dimensional (2D) artificial atoms\cite{1,2} and in order to get insight on the
behaviour of 2D systems in the quantum Hall regime. Special
attention has been given to 2D parabolic quantum dots (PQD) since the
non-interacting system is analytically solvable, its high symmetry
symplifies the calculations and some of its properties do not depend on
the strengh of the confinement potential.

In this paper we introduce the set of operators describing low-lying
excitations in PQD for filling factor $ \nu = 1$. There are two
key conditions to obtain these operators: Firstly, the fact that, close
to $\nu =1$, the lowest excitation within each total angular momentum subspace
are excitations of integer charge and, secondly, to impose to the
eigenstates these operators generate a well defined third component
of both total ($M$) and center-of-mass ($M_{CM}$) angular momentum.
At the end of the paper we discuss several applications
of such operators in the analysis and calculation of relevant magnitudes
for transport and spectroscopy in QHD.

We consider interacting electrons in a PQD affected by a perpendicular
magnetic field $B$ high enough for the N-electron wave
functions to comprise only polarised single particle states
in the lowest Landau level (LLL). Due to the circular symmetry, it is
convenient to work in the symmetric gauge, so that
$M$ is a good quantum number. The Hamiltonian can be written as:
\begin{eqnarray}
H \! = \! [\frac {N}{2}+\frac{1 - \gamma}{2}M] \hbar \Omega + \!
\sum_{\{m_i\}} \! \frac{V_{m_1m_2m_3m_4}}{2}c^{\dagger}_{m_1}
c^{\dagger}_{m_2}c_{m_3}c_{m_4} \nonumber
\end{eqnarray}
where the sum runs over the single-particle angular momentum labelling the
states of a PQD defined in the $xy$ plane, and
characterized by a frequency $\omega_0$.
$\gamma = \omega _{c} /\Omega $, $ \omega _{c} = (e B)/(m ^ \ast c)$,
$ \Omega = (\omega _{c} ^{2} + 4 \omega _0^{2} ) ^{1/2} $, and
$ \l_B = \sqrt{(\hbar)/(m ^ \ast \Omega )} $ is the effective magnetic length.
$c^{\dagger}$ and $c$ are the electron creation and annihilation operators.
An additional advantage of PQD is that the electron-electron interaction
matrix elements $ V_{m_1m_2m_3m_4}$ between states
$\{m\}$ in 2D can be computed analytically\cite{Girvinjach,Stone2}. In
this paper we take the interaction as coulombic, unless explicitely stated
otherwise. Let us review some of the known properties of this Hamiltonian.
Since the interaction only depends on the
relative motion of electrons, the QHD spectrum separates in subspaces with
different value of $M$\cite{3,4,5,10}.
Due to the reduction of the Hilbert space to the LLL,
the dimension of each subspace labelled by $(N,M)$ is finite.
Remarkably, wave functions do not depend on $ \omega_0 $ or
$B$ (apart from a rescaling of the length scale), neither does their
energy ordering $within$ a $(N,M)$ subspace. This property can be used
to get information on an infinite system from the results in a PQD.
Only the relative positioning of energies for different $(N,M)$ subspaces
depend on the values of the parameters defining the QD due to the first
term of the Hamiltonian. On increasing $B$ (or,
alternatively, decreasing $ \omega_0 $) the $N$-electron GS resides in
$(N,M)$ subspaces with increasingly larger $M$\cite{3,4,5,10}.

We denote the GS in subspace $(N,M)$ by $ \mid \Phi _{M}^{(N)}\rangle  $.
An important fact, which to the best of our knowledge, has never been stated
before, is that $\langle N_E\rangle _{M}= \langle \Phi _{M}^{(N)} \mid
\sum_{m=N}^{\infty} c^{\dagger}_{m} c_{m} \mid \Phi _{M}^{(N)}\rangle $
is extremely close to an integer. As $M$ increases,
$\langle N_E\rangle _{M}$ increases in steps of heigh 1, and size typically
$N$, up to a number of steps of the order of $N/2$ beyond which this property
fails\cite{nota1}. We have checked that this property holds not only for
the Coulomb interaction, but also for the logarithmic and contact interactions.
Therefore, in order to study the low-lying excitations close to
$M_0$, we can consider a family of variational wave functions, each one
of them being a linear combination of Slater determinants with the same
value of $\langle N_E\rangle _{M}$. This additional "symmetry" allows the
computation of properties in this system for very large values of $N$ and $M$.

In this letter we are going to concentrate on the study of the first
sector, i.e. when $\langle N_E\rangle _{M} \approx 1$. Numerically, we find
that
there is always an eigenstate with the previous property in each subspace
$(N,M)$ with $M_0 < M \leq M_0+N$. Excitations with $\langle N_E\rangle _M=1$
can be expressed as a representation of the charge density operator
projected onto the LLL.  They resemble the single-mode
excitations previously studied in superfluid $^{4}$He and
in the fractional Quantum Hall regime\cite{Girvin}. A set of such operators
which has been put forward\cite{Wen,Stone,Stone2,Marsili,MacDonald,Pitaevskii}
to describe boson-like edge-wave excitations in PQD is
\begin{eqnarray}
S^\dagger _{\Delta M}=
\sum_{k=1}^{N} (b_k^\dagger /2)^{\Delta M}=
\sum_{m} \sqrt{\frac{(m+\Delta M)!}{m!}}
c_{m+\Delta M}^\dagger c_m. \nonumber
\end{eqnarray}
$b_k^\dagger =-(x_k+iy_k)/2l_B=-z_k/2$ is, apart from a normalization constant,
the ladder operator increasing the single particle angular momentum in the LLL.
It has been conjetured\cite{Stone2}, and proven for the parabolic
electron-electron
interaction\cite{Marsili}, that these operators represent a bosonisation
of the low-lying excitations for $\Delta M <<N$. However the
overlap of the states they generate with the true ground state decreases
rapidly with $\Delta M$ being negligible for $\Delta M \approx N$ (see
below). The reason for this is that $S^\dagger _{\Delta M}\mid \Phi_{M_0}^
{(N)}\rangle $, for $\Delta M >1$, {\em is not an eigenstate of the third
component
of the center-of-mass angular momentum}. Note that $S^\dagger _{1}, $
when operated on an {\it exact} eigenstate, gives an eigenstate with
$M_{CM}$ increased by 1\cite{trugman}. States with well defined $M_{CM}$ should
be anihilated by
$(S_1)^{M_{CM}}$. It is straighforward to check that the condition
$(S_1)^{M_{CM}} S_{\Delta M}^\dagger \mid \Phi _{M_0}^{(N)}\rangle =0$ is
not fulfilled for $\Delta M>1$.

The previous discussion gives the key idea for finding boson-like excitations
(with well defined $M$ and $M_{CM}$) by means of an operator $J^\dagger
_{\Delta M}$ which ejects one electron from the compact state.
Since these excitations must have the lowest
possible energy, the center-of-mass tends to be at rest, implying the condition
$S_1 J_{\Delta M}^\dagger \mid \Phi _{M_0}^{(N)}\rangle =0$.
 From this condition we get,
\begin{eqnarray}
J^\dagger _{\Delta M}\! =\!  \sum_{m}\!  \sqrt{\frac{m!}{(m+\Delta M)!}}
c_{m+\Delta M}^\dagger c_m \! = \! \sum_{k=1}^{N} [b_k^\dagger
(b_k b_k^\dagger )^{-1}]^{\Delta M} \nonumber
\end{eqnarray}
for $1<\Delta M \leq N$. These operators can be written as
$J^\dagger _{\Delta M}=2^{\Delta M/2} \int dz (z^*)^{-\Delta M} \rho _0 (z)$.
They are the representation of the projection on the LLL of the
density operator ($\rho _0 (z)$)
on the subset ($(z^*)^{-\Delta M}$) of the cylindrical harmonics\cite{ss}.
 From the first quantisation expression for $J^\dagger _{\Delta M}$,
we find\cite{tobe} that it is possible to expand $J^\dagger _{\Delta M}$ in
a series which i) for $\Delta M=N>>1$ gives the Laughlin\cite{3}
wavefunction for a quasi-electron: $\mid \Phi _{N}^{(N)}\rangle = \prod z_i
\mid \Phi _{M_0}^{(N)}\rangle $, and ii) has $S^\dagger _{\Delta M}$ as
asymtotic limit for $\Delta M << N$. In between these two limits the
density charge excitation created by the operator $J^\dagger _{\Delta M}$
consists of a "dressed" electron-hole pair, i.e., a {\em magnetoexciton},
where the distance between the hole and the electron is proportional
to the quantum number ${\Delta M}$. The electron is always
located at the edge of
the droplet while the hole moves toward the center of the droplet with $\Delta
M$. For $\Delta M = N$, the hole reaches the centre of the droplet and
this marks the end of the first sector.

In order to check the quality of $J^\dagger _{\Delta M}$ for describing
QHD excitations, table I shows the overlaps
$\langle \Phi_{M}^{(N)}|J^\dagger _{\Delta M}|\Phi_{M_0}^{(N)}\rangle$ and
$\langle \Phi_{M}^{(N)}|S^\dagger _{\Delta M}|\Phi_{M_0}^{(N)}\rangle$
for $N=10,20$ and $\Delta M \le N$.
Note that for ${\Delta M}\leq 3$, $J^\dagger _{\Delta
M}|\Phi_{M_0}^{(N)}\rangle$
is an exact eigenstate. Using Trugman and Kivelson's \cite{trugman}
procedure to generate eigenstates it is possible to show that this is so
for any isotropic pairwise interaction.
The conclusion from table I is clear: the branch
of lowest excitations are magnetoexcitons with the center-of-mass at rest.
$S^\dagger _{\Delta M}$ only describe approximately the excitations in the
$\Delta M \ll N $ limit \cite{Stone,Stone2,Marsili,MacDonald,Pitaevskii} and,
as expected, become poor descriptions of the excitations for droplets with
small number of electrons. The operators $J^\dagger _{\Delta M}$ represent
a bosonization of low-lying excitations of the PQD close to $\nu =1$. As it
happens in other cases of bosonization\cite{mahan},
they do not satisfy bosonic commutation relations strictly; however
their expectation values on $\mid \Phi_{M_0}^{(N)}\rangle$ do.

 From the analytical expression for the wavefunctions it is possible
to compute their energies
$E_M^{(N)}=\langle \Phi_{M_0}^{(N)}|J_{\Delta M}HJ_{\Delta M}^\dagger |
\Phi_{M_0}^{(N)} \rangle / \langle \Phi_{M_0}^{(N)}|J_{\Delta M}
J_{\Delta M}^\dagger |\Phi_{M_0}^{(N)} \rangle $, for large values of $N$.
Using Wick's theorem, the energy difference between the
GS in the $(N,M)$ and $(N,M_0)$ subspaces can be written as:
\begin{eqnarray}
\hbar \Omega (1-\gamma ) \Delta M/2\! + \! \epsilon _{d-e}^{(N)}(\Delta M)\!
+ \! \epsilon _{d-h}^{(N)}(\Delta M)\! + \! \epsilon _{e-h}^{(N)}(\Delta M).
\nonumber
\end{eqnarray}
The first term is the single-particle energy difference between the two states.
The remaining three terms are the contributions coming from the
interaction; they involved combinations of interaction matrix elements,
and can be identified as: $\epsilon _{d-e}^{(N)}$ is the interaction
between the compact $\nu=1$ droplet and the ejected electron onto the edge
and is practically independent on $\Delta M$.
$\epsilon_{d-h}^{(N)}$ is the energy corresponding to the
interaction between the $\nu=1$ droplet and the hole created in it, and
decreases almost linearly with $\Delta M$.
$\epsilon_{e-h}^{(N)}$ represents the binding energy of the
magnetoexciton, and presents a minimum as a function of $M$.
Figure \ref{fig2} shows the dependence with $M$ of the these last three terms
for $N=30$. We find that the total energy difference has the magnetoroton
minimum at the same position as $\epsilon _{e-h}^{(N)}$ in the region where
the description in terms of $J^\dagger _{\Delta M}$ is valid.
The magnetoroton minimum marks the $ \Delta M$ at which the first
instability in the $\nu=1$ droplet occurs.
We find that, for a contact interaction, the position of this intra-Landau
magnetoroton at $\nu =1$ shifts as $2\sqrt {N}$ in the whole range of $N$.

We consider now  another important application of the operators
$J^\dagger_{\Delta M}$. Several transport and spectroscopic properties
\cite{7,beena,epl} are controlled, close to $\nu =1$, by the spectral function
\begin{eqnarray}
A(N,\omega )=\! \sum _{\eta ,m}\!
|\langle \Phi_\eta ^{(N)} | c_{m}^\dagger | \Phi_{M_0} ^{(N-1)} \rangle |^2
\! \delta (\omega - \! E_\eta ^{(N)}+\! E_0 ^{(N-1)}) \nonumber
\end{eqnarray}
where $\eta $ labels all $N$-particle states.
In spite of the fact that magnetoexcitons $J^\dagger _{\Delta M}
|\Phi_{M_0}^{(N)}\rangle$ are not the only excitations in the system,
the main contributions to $A(N,\omega )$ come from states of
the form $(S^\dagger _1)^j J^\dagger _{\Delta M} \mid \Phi_{M_0} ^{(N)}\rangle
$,
that is, from magnetoexcitons and all the states generated from them
by increasing $M_{CM}$. Figure \ref{fig3} shows the
comparison between the exact spectral function $A(N,\omega )$ (lower panel)
and the one calculated including only the preceeding states (upper panel),
for $N=8$ and at a value of $\Omega $ for which the GS of the
7-electron PQD is the compact state.
The identification of the relevant states for the calculation of
$A(N,\omega )$ allows i) the analytical calculation of this magnitud for
large N and ii) the understanding of the different bunches of peaks spaced in
$(\Omega - \omega _c)/2 $ and decreasing in intensity that appear in
$A(N,\omega )$.  After the first high peak coming from the transition
between compact states in both $N-1$ and $N$-particle systems, the first bunch
is due to the transition from the compact
state of $N-1$ particles to magnetoexcitons of $N$ particles.
Each subsequent bunch of peaks in $A(N,\omega )$ is a replica of the
preceeding one with $M_{CM}$ increased in 1.

Let us now pay attention to the spectrum for large values of
$\Delta M$. The lowest-lying states of sectors beyond the first, and even
at the end of the first sector ($\Delta M \approx N $)
for large values of $N$, cannot be described in terms of single
magnetoexcitons. This is observed in table I where the overlap at
$\Delta M=N=20$ is small\cite{nota3}.
Since an integer number of electrons is ejected from the compact state, one
is tempted to look for almost independent bosons in which
low-lying states are obtained by succesive application of $J^\dagger
_{\Delta M}$. Comparing the energies so
obtained with the exact ones, the result is satisfactory.
However this is not generally the case when comparing wavefunctions. The state
$J^\dagger _{\Delta M_1} J^\dagger _{\Delta M_2}
|\Phi _{M_0} ^{(N)}\rangle $ with $\Delta M_1+\Delta M_2>N$
has a small overlap with the state $|\Phi
^{(N)}_{M_0+\Delta M_1+\Delta M_2}\rangle $ obtained by exact diagonalization.
Only when $\Delta M_1=\Delta M_2=N$, the overlap is close to 1 and the state
can be considered as constituted by two almost non-interacting magnetoexcitons.

Finally, let us discuss one implication of the increasing correlation
effects with increasing $B$.
GS changes from the compact state to a magnetoexciton
and further to states with several holes in the compact and the same
number of electrons out of it.
Once spectral functions are known, single electron capacitance
experiments in QD\cite{2} can be analized.
For low temperature, tunneling rates depend only on the
spectral weight $\Delta (N)=A(N,\mu (N))$\cite{nos}. This magnitude is
shown in fig. \ref{fig4} as a function of $B$. For a field
in which the GS of both $N-1$ and $N$ electron systems are the compact
ones, the spectral weight is obviously 1. For increasing field those two
GS melt by ejecting 1 electron. As holes produced in the inner part
of the dot are not completely equivalent, $\Delta (N)$ decreases.
In some cases the value of the field for which melting appears is
significantly different between $N-1$ and $N$ particles so that
the spectral weight is drastically quenched. The reduction of
$\Delta (N)$ with $B$ explains the experimental quenching of
signal amplitude in single electron capacitance experiments for high
magnetic fields\cite{2}.

In summary, we have found the adequate circular representation of charge
density excitations in QHD in $\nu =1$. Magnetoexcitons
$J^\dagger _{\Delta M}|\Phi_{M_0}^{(N)}\rangle$ turn out to be an excellent
description of low-lying excitations. From such wavefunctions it is
easy to get energies and spectral functions, i. e. the important
information on the electronic structure of QHD.

We thank L. Brey for many useful discussions.
This work has been supported in part by the Comisi\'on Interministerial
de Ciencia y Tecnolog\'{\i}a of Spain under contracts No. MAT 91-0905 and
MAT 94-0982-C02-01 and by the Commission of the European Communities under
contract No. SSC-CT-90-0020.

\begin{table}
\caption{Overlaps between the exact lowest eigenstate $\Phi_{M}
(N)$ and the states obtained applying respectively $J^\dagger _{\Delta M}$ and
$S^\dagger _{\Delta M}$ to the compact GS for $N=10$ and $20$. The small
overlap
$\langle \Phi_{M_0+20}^{(20)}|J^\dagger _{20}|\Phi_{M_0}^{(20)}\rangle$
is discussed in the text.}

\begin{tabular}{|c|cccc|}

& \multicolumn{2}{c}{$N=10$} & \multicolumn{2}{c|}{$N=20$} \\ \hline
$ \Delta M $
&$J^\dagger _{\Delta M}$ &$S^\dagger _{\Delta M}$ &$J^\dagger _{\Delta M}$
&$S^\dagger _{\Delta M}$
\\ \hline
1 & 1.000000 & 1.000000 & 1.000000 & 1.000000 \\
2 & 1.000000  & 0.994988  &1.000000  & 0.998749 \\
3 & 1.000000 &  0.970047   &1.000000  & 0.992503 \\
4 &0.999906  & 0.906678 &0.999960  & 0.976401 \\
5 &0.999589&  0.788718   &0.999752  & 0.944870 \\
6 &  0.999285&   0.614055 &0.999220  & 0.892297 \\
7 &  0.999378 &  0.407942 &0.998367  & 0.814360 \\
8 &  0.999668 &  0.219840 & 0.997480 &  0.710019 \\
9 &  0.999831 &  0.090603 &0.992292  & 0.482037\\
10 &  0.999923 &  0.024995 &0.993332  & 0.315831 \\
11 &                 &           &0.995374  & 0.196493 \\
12 &         &        &0.996878  & 0.114977 \\
13 &         &        &0.997760  & 0.082670\\
14 &        &         &0.998176  & 0.057225 \\
15 &        &        &0.998507  & 0.037717 \\
16 &      &     & 0.998866  & 0.023329 \\
17 &      &     & 0.999255  & 0.013203 \\
18 &      &     &  0.999537 &  0.006665 \\
19 &      &     & 0.998329  & 0.003121 \\
20 &      &     & 0.000222  & 0.000921 \\
\end{tabular}
\label{table}
\end{table}

\begin{figure}
\caption{Different contributions to the magnetoexciton energy as a function of
$\Delta M$ for $30$ particles.}
\label{fig2}
\end{figure}

\begin{figure}
\caption{(a) Spectral function for 8 electrons including only magnetoexcitons
with
rotating center-of-mass ($(S^\dagger _1)^j J^\dagger _{\Delta M}|\Phi_{M_0}
^{(N)}\rangle $). (b) Total spectral function including the whole
spectrum obtained from
the exact diagonalization. }
\label{fig3}
\end{figure}

\begin{figure}
\caption{Many-body contribution to the low-temperature
tunneling rates of an electron entering in a QD
as given by the spectral weight $\Delta (N)$ as a function of $B$ for
different number of electrons.}
\label{fig4}
\end{figure}


\begin{references}
\bibitem{nota1}For instance, for $N=8$, $\langle N_E\rangle _M$ in the first
sector is 1 while in the second is $2>\langle N_E\rangle _M>1.95$ and
$3>\langle
N_E\rangle _M>2.88$ in the third sector.
\bibitem{ss}$S^\dagger _{\Delta M} = 2^{-\Delta M/2} \int dz z^{\Delta
M} \rho _0 (z)$ are a representation of $\rho _0$ on other
subset ($z^{\Delta M}$) of the cylindrical harmonics.
\bibitem{nota3}An overlap of 0.9982 between the magnetoexciton and the
first exact excited state is obtained.
\end{references}
\end{document}